\documentstyle[11pt]{article}
\begin{document}

\title{\textbf{Cosmological constraints on polytropic gas model}}

\author{K. Karami\thanks{E-mail: KKarami@uok.ac.ir} , Z. Safari, S.
Asadzadeh\\\\
\small{Department of Physics, University of Kurdistan, Pasdaran St.,
Sanandaj, Iran}}

\maketitle

\begin{abstract}
We study the polytropic gas scenario as the unification of dark
matter and dark energy. We fit the model parameters by using the
latest observational data including type Ia supernovae, baryon
acoustic oscillation, cosmic microwave background, and Hubble
parameter data. At 68.3\% and 95.4\% confidence levels, we find the
best fit values of the model parameters as
$\tilde{K}=0.742_{-0.024}^{+0.024}(1\sigma)_{-0.049}^{+0.048}(2\sigma)$
and $n=-1.05_{-0.08}^{+0.08}(1\sigma)_{-0.16}^{+0.15}(2\sigma)$.
Using the best fit values of the model, we obtain the evolutionary
behaviors of the equation of state parameters of the polytropic gas
model and dark energy, the deceleration parameter of the universe as
well as the dimensionless density parameters of dark matter and dark
energy. We conclude that in this model, the universe starts from the
matter dominated epoch and approaches a de Sitter phase at late
times, as expected. Also the universe begins to accelerate at
redshift $z_{\rm t}=0.74$. Furthermore in contrary to the
$\Lambda$CDM model, the cosmic coincidence problem is solved
naturally in the polytropic gas scenario.
\end{abstract}


\clearpage

\section{Introduction}
Various cosmological observations, including the type Ia supernovae
(SNeIa) \cite{Perlmutter}, cosmic microwave background (CMB)
\cite{Bennett} and baryon acoustic oscillation (BAO)
\cite{Eisenstein}, etc., have revealed that the current expansion of
the universe is accelerating and it entered this accelerating phase
only in the near past. A new energy component with negative pressure
called dark energy (DE) is needed to explain this acceleration
expansion within the framework of general relativity (GR). The
simplest and most appealing candidate for DE is the $\Lambda$ cold
dark matter ($\Lambda$CDM) model, in which the cosmological constant
$\Lambda$ plays a role of DE in GR. This model is in general
agreement with current astronomical observations, but has
difficulties in reconciling the small observational value of DE
density to that coming from quantum field theories. This is called
the cosmological constant problem \cite{Weinberg}. Therefore, a
number of models for DE to explain the current acceleration without
the cosmological constant has been proposed, such as quintessence
\cite{8}, phantom \cite{9}, K-essence \cite{10}, tachyon \cite{11},
quintom \cite{12}; as well as the Chaplygin gas \cite{13} and the
generalized Chaplygin gas \cite{14}, the holographic DE \cite{15},
the new agegraphic DE \cite{16}, the Ricci DE \cite{17}, and so on.

One of interesting DE models is the polytropic gas which was
proposed to explain the accelerated expansion of the universe
\cite{Mukhopadhyay,Karami1,Christensen}. It was shown that the
polytropic gas model in the presence of interaction with DM can
behave as phantom type DE \cite{Karami1}. It was pointed out that a
polytropic scalar field can be reconstructed according to the
evolutionary behaviors of the holographic \cite{Karami2} and new
agegraphic \cite{Karami3} DE densities. The validity of the
generalized second law of gravitational thermodynamics was examined
for the polytropic gas model in \cite{Karami4}. Different modified
gravity models including $f(T)$-gravity \cite{Karami5} and
$f(R)$-gravity \cite{Karami6} were reconstructed according to the
polytropic gas equation of state. The polytropic tachyon, K-essence
and dilaton scalar field models were studied in \cite{Malekjani1}.
In \cite{Malekjani2}, the polytropic gas model was investigated from
the viewpoint of statefinder diagnostic tool and $w-w'$ analysis.

In the present work, our main aim is to fit the polytropic gas model
and give the constraints on model parameters, with current
observational data including SNeIa, CMB, BAO and observational
Hubble data (OHD), filling in the gap existing in the literature. To
do so, in section 2, we briefly review the polytropic gas model as
unified picture of DM and DE in a spatially flat
Friedmann-Robertson-Walker (FRW) universe. In section 3, we consider
the cosmological constraints on the polytropic gas model by using
the latest observational data. In section 4, we give numerical
results. Section 5 is devoted to conclusions.

\section{Polytropic gas model}

For a polytropic gas model, the energy density $\rho_{\rm pol}$ and
pressure $p_{\rm pol}$ satisfy the equation of state (EoS)
\cite{Christensen}
\begin{equation}\label{EoSpol}
p_{\rm pol}=-K\rho_{\rm pol}^{1+\frac{1}{n}},
\end{equation}
where $K>0$ and the polytropic index $n$ are two constants of the
model.

Here, we consider a spatially flat FRW universe containing only the
polytropic fluid and baryonic matter. We ignore the contribution of
radiation, as expected from the observations \cite{Komatsu}. In the
framework of the standard FRW cosmology, the first Friedmann
equation reads
\begin{equation}\label{Feq}
    H^{2}=\frac{8\pi G}{3}(\rho_{\rm pol}+\rho_{\rm bm}),
    \end{equation}
where $H=\dot{a}/a$ is the Hubble parameter. Also $\rho_{\rm pol}$
and $\rho_{\rm bm}$ denote the energy densities of the polytropic
and pressureless baryonic matter ($p_{\rm bm}=0$), respectively,
which satisfy the following energy conservation laws
\begin{equation}\label{Eeqpol}
        \dot{\rho}_{\rm pol}+3H(\rho_{\rm pol}+p_{\rm pol})=0,
    \end{equation}
    \begin{equation}\label{Eeqbm}
        \dot{\rho}_{\rm bm}+3H\rho_{\rm bm}= 0.
    \end{equation}
Using the dimensionless density parameters
\begin{equation}\label{OmegaP}
\Omega_{\rm pol}=\frac{\rho_{\rm pol}}{\rho_{\rm cr}}= \frac{8\pi
G\rho_{\rm pol}}{3H^{2}},~~~\Omega_{\rm bm}= \frac{\rho_{\rm
bm}}{\rho_{\rm cr}}= \frac{8\pi G\rho_{\rm bm}}{3H^{2}},
\end{equation}
the Friedmann equation (\ref{Feq}) is rewritten as follows
\begin{equation}\label{Feq2}
1=\Omega_{\rm pol}+\Omega_{\rm bm}.
\end{equation}
Substituting Eq. (\ref{EoSpol}) into (\ref{Feq}) gives the evolution
of the polytropic gas energy density in terms of redshift
$z=\frac{1}{a}-1$ as
\begin{equation}\label{rhopol}
\rho_{\rm pol}=\rho_{\rm pol_0}\Big[
\tilde{K}+(1-\tilde{K})(1+z)^{-3/n}\Big]^{-n},
\end{equation}
where
\begin{equation}
\tilde{K}=\rho_{\rm pol_0}^{1/n}K,
\end{equation}
and $\rm{\rho_{\rm pol_0}}$ is the polytropic energy density at
present time. Also Eq. (\ref{Eeqbm}) gives
\begin{equation}
\rho_{\rm bm}=\rho_{\rm bm_0}(1+z)^3.\label{rhobm}
\end{equation}
It is worth to note that the polytropic gas energy density
(\ref{rhopol}) offers a unified picture of DM and DE. Because it
smoothly interpolates between a non-relativistic matter phase,
$\rho_{\rm pol}\propto (1+z)^3$, in the past and a negative-pressure
DE regime, $\rho_{\rm pol}=-p_{\rm pol}$, at late time
($z\rightarrow-1$). Therefore, within the framework of unified DM-DE
(UDME) scenario one can rewrite the polytropic gas energy density as
\begin{eqnarray}
&&\rho_{\rm pol}=\rho_{\rm dm}+\rho_{\rm de},\nonumber\\
&&p_{\rm pol}=p_{\rm de},\label{rho-p-pol}
\end{eqnarray}
where the DM is a pressureless matter ($p_{\rm dm}=0$). Using Eqs.
(\ref{EoSpol}) and (\ref{rhopol}) one can obtain the EoS parameter
$\omega_{\rm pol}$ of the polytropic gas model as
\begin{equation}\label{EOS}
\omega_{\rm pol}(z)=\frac{p_{\rm pol}}{\rho_{\rm
pol}}=-\frac{\tilde{K}}{\tilde{K}+(1-\tilde{K})\left(1+z\right)^{-3/n}}.
\end{equation}
With the help of Eqs. (\ref{Eeqpol}), (\ref{Eeqbm}), (\ref{rhopol})
and (\ref{rho-p-pol}), the energy densities of DM and DE evolve as
follows
\begin{equation}\label{rho de}
\rho_{\rm dm}=\rho_{\rm dm_0}(1+z)^{3},
\end{equation}
\begin{equation}\label{rho de}
\rho_{\rm de}=\rho_{\rm
pol_0}\Big[\tilde{K}+(1-\tilde{K})(1+z)^{-3/n}\Big]^{-n}
 -\rho_{\rm dm_0}(1+z)^{3}.
\end{equation}
Using Eqs. (\ref{EoSpol}), (\ref{OmegaP}) and (\ref{rho de}) the EoS
parameter of DE reads
\begin{equation}
\omega_{\rm de}(z)=\frac{p_{\rm de}}{\rho_{\rm
de}}=\frac{-\tilde{K}(1-\Omega_{\rm
bm_{0}})\Big[\tilde{K}+(1-\tilde{K})(1+z)^{-3/n}\Big]^{-1-n}}{(1-\Omega_{\rm
bm_0})\Big[\tilde{K}+(1-\tilde{K})(1+z)^{-3/n}\Big]^{-n}-\Omega_{\rm
dm_0}(1+z)^3}.\label{wde}
\end{equation}
With the help of Eqs. (\ref{rhopol}) and (\ref{rhobm}) and using
(\ref{OmegaP}), the first Friedmann equation (\ref{Feq}) takes the
form
\begin{equation}\label{H}
H^{2}(z,\textbf{p})=H_{0}^{2}\left\{(1-\Omega_{\rm
bm_0})\Big[\tilde{K}+(1-\tilde{K})(1+z)^{-3/n}\Big]^{-n}+\Omega_{\rm
bm_0}(1+z)^{3}\right\},
\end{equation}
where $H_0 = 70.2\pm 1.4~\rm km~s^{-1}~Mpc^{-1}~(68\%~\rm CL)$ and
$\Omega_{\rm bm_0}=\frac{8\pi G\rho_{\rm bm_0}}{3H_0^2}=0.0458\pm
0.0016~(68\%~\rm CL)$ are the present Hubble constant and the
present value of the dimensionless BM density, respectively, which
have been updated in the 7-year WMAP (WMAP7) data \cite{Komatsu}.
Also \textbf{p} indicate model parameters including $\tilde{K}$ and
$n$. Thus, throughout this work we fix the Hubble and the baryon
density parameters at $H_0 = 70.2$ and $\Omega_{\rm bm_0}= 0.0458$.
With $H_0$ and $\Omega_{\rm bm_0}$ being determined by independent
measurements, in the next section we will use the cosmic
observations to constrain the polytropic gas model parameters
($\tilde{K},n$).

For completeness, we give the deceleration parameter
\begin{equation}
q=-1-\frac{\dot{H}}{H^2},\label{qdif}
\end{equation}
which combined with the EoS and the dimensionless density parameters
form a set of useful parameters for the description of the
astrophysical observations. Using Eq. (\ref{H}) the deceleration
parameter (\ref{qdif}) can be obtained as
\begin{equation}
q(z)=\frac{1}{2}\frac{\Omega_{\rm bm_{0}}(1+z)^{3}+(3\omega_{\rm
pol}+1)(1-\Omega_{\rm
bm_{0}})\left[\tilde{K}+(1-\tilde{K})(1+z)^{-3/n}
\right]^{-n}}{\Omega_{\rm bm_{0}}(1+z)^{3}+(1-\Omega_{\rm
bm_{0}})\left[\tilde{K}+(1-\tilde{K})(1+z)^{-3/n}
\right]^{-n}}.\label{qdec}
\end{equation}

\section{Observational constraints}

Here, we fit the free parameters of the polytropic gas model
(\ref{EoSpol}) by using the recent observational data including
SNeIa, BAO, CMB and OHD.

For the SNeIa data, we use the currently largest Union2.1
compilation \cite{union} that contains a total of 580 SNeIa, which
is an updated version of the Union2 compilation \cite{Amanullah}.
Cosmological constraints from SNeIa data are obtained through the
distance modulus $\mu(z)$. The theoretical distance modulus is
defined as \cite{Pietro,Nesseris}
\begin{equation}
\mu_{\rm th}(z)=5\log_{10}D_{\rm L}(z)+\mu_{0},
\end{equation}
where $\mu_{0}=42.38-5\log_{10}h$ and $h$ is the Hubble constant
$H_0$ in units of $100~\rm{km~s^{-1}~Mpc^{-1}}$. Also the
Hubble-free luminosity distance $D_{\rm L}(z)$ for the flat universe
is given by
\begin{equation}
 D_{\rm L}(z)=(1+z)\int_{0}^{z}\frac{dz'}{E(z';\textbf{p})},\label{DL-SNeIa}
 \end{equation}
with \textbf{p} the model parameters and
$E(z;\textbf{p})=H(z;\textbf{p})/H_0$.

For using SNeIa data, theoretical model parameters are determined by
minimizing the quantity \cite{Pietro,Nesseris}
\begin{equation}
\tilde{\chi}_{\rm SN}^{2}=A-\frac{B^{2}}{C},
\end{equation}
where
\begin{equation}
A=\sum_{\rm i=1}^{580}[\mu_{\rm obs}(z_{\rm i})-\mu_{\rm th}(z_{\rm
i})]^{2}/\sigma_{\rm i}^{2},
\end{equation}
\begin{equation}
B=\sum_{\rm i=1}^{580}[\mu_{\rm obs}(z_{\rm i})-\mu_{\rm th}(z_{\rm
i})]/\sigma_{\rm i}^{2},
\end{equation}
\begin{equation}
C=\sum_{\rm i=1}^{580}1/\sigma_{\rm i}^{2},
\end{equation}
and $\sigma_{\rm i}$ stands for the $1\sigma$ uncertainty associated
to the $i$th data point.

Next, we add the data from the observation of acoustic signatures in
the large scale clustering of galaxies. Using the BAO data, one can
minimize the $\chi_{\rm BAO}^{2}$ defined as
\cite{Tegmark,Eisenstein},
\begin{equation}
\chi_{\rm BAO}^{2}=\frac{\left[A_{\rm obs}-A_{\rm
th}\right]^{2}}{\sigma_{A}^{2}},
\end{equation}
where
\begin{equation}
A_{\rm th}=\sqrt{\Omega_{\rm m_0}}~E(z_{\rm
b};\textbf{p})^{-1/3}\left[\frac{1}{z_{\rm b}}\int_{0}^{z_{\rm
b}}\frac{dz'}{E(z';\textbf{p})}\right]^{2/3},\label{A-BAO}
\end{equation}
is the theoretical distance parameter and $z_{\rm b}=0.35$ is the
redshift of luminous red galaxies sample of the Sloan Digital Sky
Survey (SDSS). Here $A_{\rm obs}=0.469(n_{s}/0.98)^{-0.35}\pm 0.017$
is measured from the SDSS data \cite{Eisenstein} and the scalar
spectral index $n_{s}$ is taken to be 0.968 \cite{Komatsu}.

Since the SNeIa and BAO data contain information about the universe
at relatively low redshifts, we will include the CMB shift
information by using the WMAP7 data \cite{Komatsu} to probe the
entire expansion history up to the last scattering surface. The
shift parameter $R$ of the CMB is defined as \cite{YWang,Bond}
\begin{equation}
R_{\rm th}=\sqrt{\Omega_{\rm m_0}}\int_{0}^{z_{\rm
rec}}\frac{dz'}{E(z';\textbf{p})},\label{R-CMB}
\end{equation}
where $z_{\rm rec}\simeq1091.3$ is the redshift at the recombination
epoch \cite{Komatsu}. Also $\Omega_{\rm m_0}$ is the effective
matter density parameter defined as
\begin{equation}
\Omega_{\rm m_0}=\Omega_{\rm bm_0}+(1-\Omega_{\rm
bm_0})(1-\tilde{K})^{-n}.\label{Omegaeff}
\end{equation}
This expression for $\Omega_{\rm m_0}$, is an estimate of the
``matter'' component of the polytropic gas fluid with the baryon
density. The $\chi^2$ from the CMB constraint is given by
\begin{equation}
\chi_{\rm CMB}^{2}=\frac{\left[R_{\rm obs}-R_{\rm
th}\right]^{2}}{\sigma_{R}^{2}},
\end{equation}
where the observational value of $R_{\rm obs}$ has been updated to
$1.725\pm0.018$ from the WMAP7 data \cite{Komatsu}.

Finally, we further add the data from the observational Hubble
parameter. The Hubble parameter is related to redshift with
\begin{equation}
H(z)=-\frac{1}{1+z}\frac{{\rm d}z}{{\rm d}t},
\end{equation}
so if ${\rm d}z/{\rm d}t$ is known, $H(z)$ is obtained directly
\cite{Hz}. The best fit values of the model parameters
 from OHD can be determined by minimizing \cite{OHD}
\begin{equation}
\chi_{\rm OHD}^{2}=\sum_{\rm i=1}^{12}\frac{\left[H_{\rm obs}(z_{\rm
i})-H_{\rm th}(z_{\rm i},\textbf{p})\right]^{2}}{\sigma_{\rm
i}^{2}},
\end{equation}
where $H_{\rm th}$ is the theoretical value of the Hubble parameter
and $H_{\rm obs}$ is the observed value. Table \ref{Hdata} shows the
observational Hubble data containing the nine data from
\cite{Hdata12} and three additional data, in bold face, from
\cite{Hdata3}.

As the relative likelihood function is defined by ${\mathcal
L}=e^{-(\chi^2_{\rm total}-\chi^2_{\rm min})/2}$ \cite{nesseris},
the best fit value of the model parameters follows from minimizing
the sum
\begin{equation}
\chi_{\rm total}^{2}=\tilde{\chi}_{\rm SN}^{2}+\chi_{\rm
BAO}^{2}+\chi_{\rm CMB}^{2}+\chi_{\rm OHD}^{2}.
\end{equation}

\section{Numerical results}

\begin{figure}
\centering\includegraphics{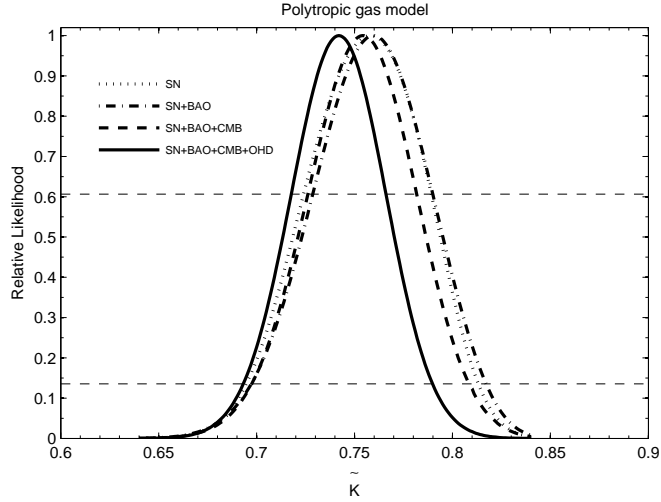}
      \vspace{5.5cm}
\caption[]{The 1D marginalized likelihood of $\tilde{K}$. The
horizontal dashed lines give the bounds with $1\sigma$ and $2\sigma$
CLs.}
         \label{1D-K}
  \end{figure}

\begin{figure}
\centering\includegraphics{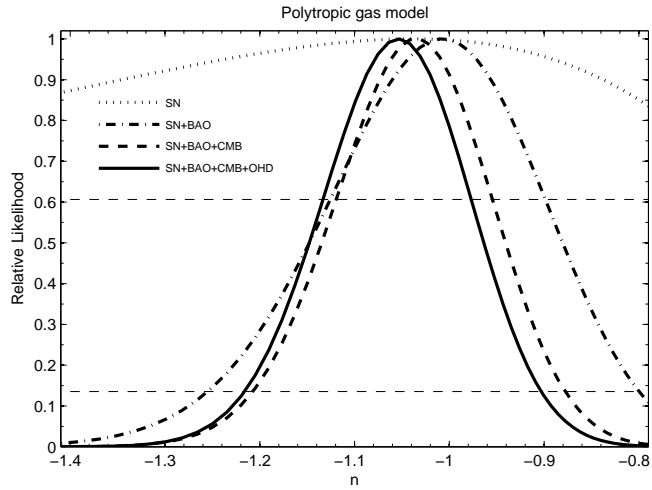}
      \vspace{5.5cm}
\caption[]{Same as Fig. \ref{1D-K}, for parameter $n$.}
         \label{1D-n}
  \end{figure}

\begin{figure}
\includegraphics{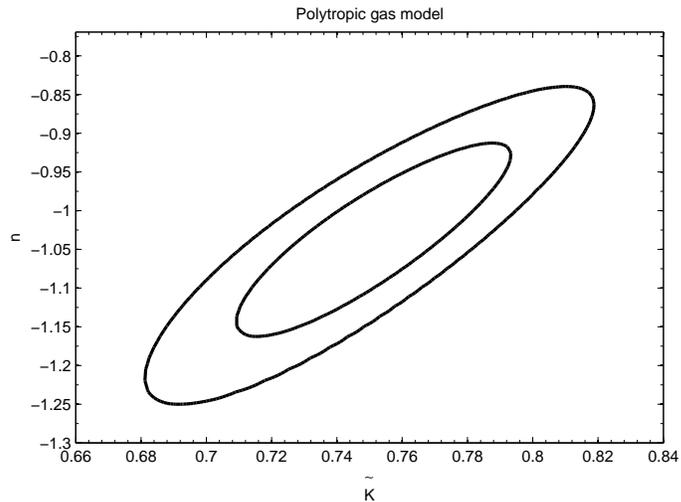}
      \vspace{5.5cm}
\caption[]{The $68.3\%~(1\sigma)$ and $95.4\% ~(2\sigma)$ CL
contours for $n$ versus $\tilde{K}$ from the full data sets.}
         \label{2D_Likelihood}
   \end{figure}

The best fit values and errors of the polytropic gas model
parameters ($\tilde{K},n$) are summarized in Table \ref{best-fit},
where we also list the best fit results of the $\Lambda$CDM model
for comparison. At $1\sigma~(68.3\%)$ and $2\sigma~(95.4\%)$
confidence levels (CLs), we obtain the best fit values
$\tilde{K}=0.742_{-0.024}^{+0.024}(1\sigma)_{-0.049}^{+0.048}(2\sigma)$
and $n=-1.05_{-0.08}^{+0.08}(1\sigma)_{-0.16}^{+0.15}(2\sigma)$ for
the full data sets including SNeIa+BAO+CMB+OHD. The total $\chi^2$
of the best fit values of the polytropic gas model is $\chi_{\rm
min}^{2}=573.089$ for the full data sets with $dof~(\rm
degree~of~freedom)=594$. The reduced $\chi^2=\chi_{\rm min}^{2}/dof$
is $0.965$, which is acceptable. The obtained $\chi_{\rm min}^{2}$
is slightly smaller than the one for the $\Lambda$CDM model,
$\chi_{\Lambda\rm CDM}^{2}=573.552$, for the same data sets. The
marginalized relative likelihood functions ${\mathcal L}(\tilde{K})$
and ${\mathcal L}(n)$ are shown in Figs. \ref{1D-K} and \ref{1D-n},
respectively. Considering only the SNeIa data, the degeneracy
between the model parameters is considerable, but using the combined
data this degeneracy is dropped in different ranges of redshift.
Figure \ref{2D_Likelihood} shows the constraint on the polytropic
gas parameter space $n-\tilde{K}$ at $1\sigma$ and $2\sigma$ CLs,
using the full data sets.

\begin{figure}
\includegraphics{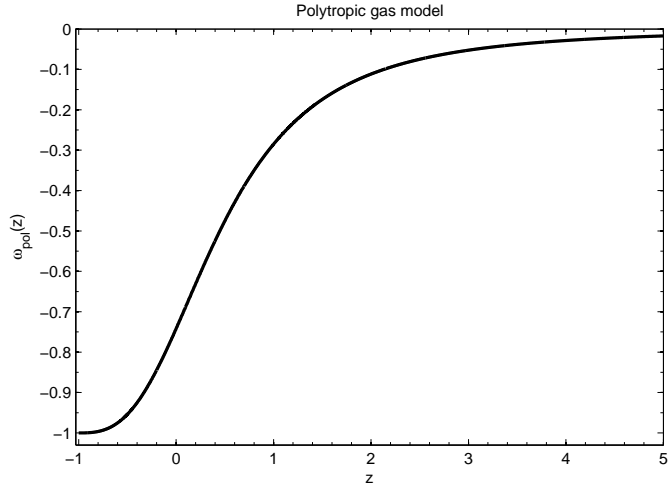}
      \vspace{5.5cm}
\caption[]{The best fit of the EoS parameter $\omega_{\rm pol}$ of
the polytropic gas model, Eq. (\ref{EOS}), using the full data
sets.}
         \label{w_PG}
   \end{figure}

\begin{figure}
\includegraphics{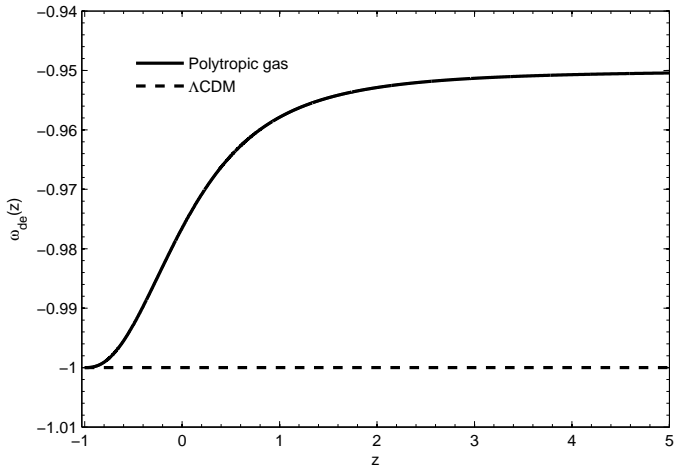}
      \vspace{5.5cm}
\caption[]{The best fit of the EoS parameter of DE $\omega_{\rm
de}$, Eq. (\ref{wde}), and the $\Lambda$CDM model using the full
data sets.}
         \label{w_de}
   \end{figure}

The variations of the EoS parameters of the polytropic gas model
$\omega_{\rm pol}$ and DE $\omega_{\rm de}$ with the best fit values
of the model are plotted in Figs. \ref{w_PG} and \ref{w_de},
respectively. Figure \ref{w_PG} presents that $\omega_{\rm pol}$ at
early ($z>>1$) and late ($z\rightarrow -1$) times behave like the
EoS parameters of the matter ($\omega_{\rm pol}=0$) and the
cosmological constant ($\omega_{\rm pol}=-1$), respectively. This
shows that in the polytropic gas scenario, the DM and DE can be
unified. Figure \ref{w_de} shows that $\omega_{\rm de}$ varies from
$\omega_{\rm de}>-1$ to $\omega_{\rm de}=-1$, which is similar to
the freezing quintessence model \cite{Caldwell2}. The current best
fit value of the EoS parameter of the DE in the polytropic gas model
is obtained as $\omega_{\rm de_0}=-0.98$ which is in good agreement
with the recent observational result $\omega_{\rm de_0}=-0.93\pm
0.13~(68\%~\rm CL)$ deduced from the WMAP7 data \cite{Komatsu}.

\begin{figure}
\includegraphics{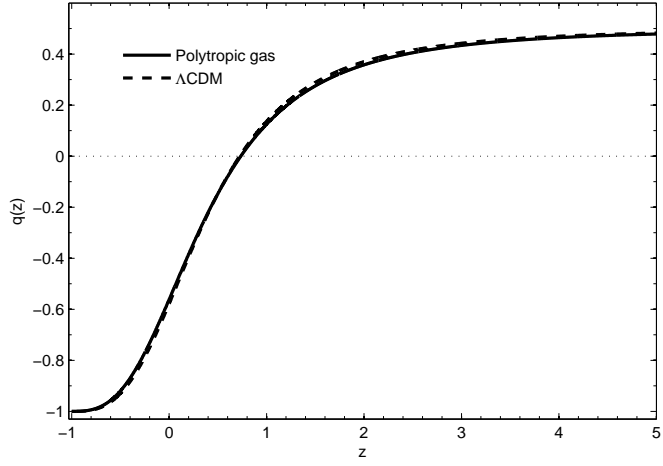}
      \vspace{5.5cm}
\caption[]{The best fit of the deceleration parameter $q(z)$ of the
universe for the polytropic gas model, Eq. (\ref{qdec}), and the
$\Lambda$CDM model using the full data sets.}
         \label{q}
   \end{figure}

In Fig. \ref{q}, we plot the evolutionary behavior of the
deceleration parameter of the universe with the best fit values of
the polytropic gas model, Eq. (\ref{qdec}), and the $\Lambda$CDM
model. Figure \ref{q} shows that very similar to the $\Lambda$CDM
model the universe transits from an early matter dominant regime,
i.e. $q=0.5$, to the de Sitter phase, i.e. $q=-1$,  in the future,
as expected. The accelerating expansion begins at transition
redshift $z_{\rm t}=0.74$, which is earlier than what the
$\Lambda$CDM model predicts, $z_{\rm t}^{\Lambda\rm CDM}=0.72$. The
current best fit value of the deceleration parameter in the
polytropic gas model is obtained as $q_0=-0.56$ which indicates the
expansion rhythm of the current universe. This is in good agreement
with the recent observational constraint
$q_0=-0.43_{-0.17}^{+0.13}~(68\%~\rm CL)$ obtained by the
cosmography \cite{CapCosmo}.

\begin{figure}
\includegraphics{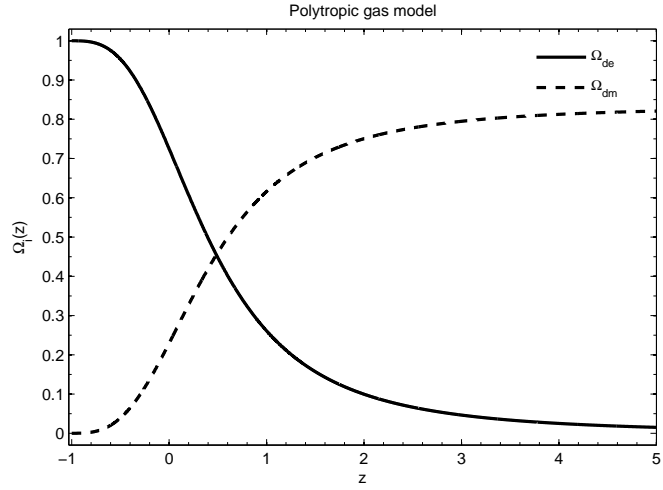}
      \vspace{5.5cm}
\caption[]{The best fits of the dimensionless density parameters of
DM, $\Omega_{\rm dm}=\frac{8\pi G\rho_{\rm dm}}{3H^2}$, and DE,
$\Omega_{\rm de}=\frac{8\pi G\rho_{\rm de}}{3H^2}$, versus redshift
for the polytropic gas model using the full data sets.}
         \label{Omega_de_dm}
   \end{figure}

In Fig. \ref{Omega_de_dm}, we plot the evolutionary behaviors of the
energy density parameters of DM, $\Omega_{\rm dm}=\frac{8\pi
G\rho_{\rm dm}}{3H^2}$, and DE, $\Omega_{\rm de}=\frac{8\pi
G\rho_{\rm de}}{3H^2}$, with the best fit values of the polytropic
gas model using the full data sets. Figure \ref{Omega_de_dm} shows
that $\Omega_{\rm dm}$ and $\Omega_{\rm de}$ decreases and
increases, respectively, during history of the universe. We also
obtain $\Omega_{\rm dm_0}=0.229$ and $\Omega_{\rm de_0}=0.725$ as
the current best fit values. These are in exact agreement with the
latest observational results $\Omega_{\rm dm_0}=0.229\pm
0.015~(68\%~\rm CL)$ and $\Omega_{\rm de_0}=0.725\pm 0.016~(68\%~\rm
CL)$ deduced from the WMAP7 data \cite{Komatsu}. We also get
$\Omega_{\rm de_0}/\Omega_{\rm dm_0}=3.166\simeq{\mathcal O}(1)$.
This shows that in contrary to the $\Lambda$CDM model, the cosmic
coincidence problem, namely why the ratio of the DE and DM densities
is of order unity today, is solved naturally in the polytropic gas
scenario.

\section{Conclusions}
Using the latest observational data from SNeIa, BAO, CMB and OHD, we
fitted the parameters of the polytropic gas model as the unification
of DM and DE. We obtained the constraint results of polytropic gas
model parameters,
$\tilde{K}=0.742_{-0.024}^{+0.024}(1\sigma)_{-0.049}^{+0.048}(2\sigma)$
and $n=-1.05_{-0.08}^{+0.08}(1\sigma)_{-0.16}^{+0.15}(2\sigma)$ for
the full data sets. The minimal $\chi^2$ gives $\chi_{\rm
min}^{2}=573.089$ with $dof=594$. The reduced $\chi^2$ equals to
$0.965$ which is acceptable. The $\chi_{\rm min}^{2}$ is slightly
smaller than the one for the $\Lambda$CDM model, $\chi_{\Lambda\rm
CDM}^{2}=573.552$, for the same data sets.

Using the best fit values of the polytropic gas model parameters, we
also studied the evolutionary behaviors of the EoS parameters of
polytropic gas model and DE, the deceleration parameter of the
universe as well as the DM and DE dimensionless density parameters.
Our numerical results show the following.

(i) The evolutionary behavior of the EoS parameter $\omega_{\rm
pol}$ of the polytropic gas model shows that the universe transits
from an early matter dominated phase, i.e. $\omega_{\rm pol}=0$, to
the $\Lambda$CDM model, i.e. $\omega_{\rm pol}=-1$, in the future,
as expected. This confirms that the polytropic gas model plays the
role of a unified model for DM and DE.

(ii) The EoS parameter $\omega_{\rm de}$ of DE in the polytropic gas
scenario varies from $\omega_{\rm de}>-1$ to $\omega_{\rm de}=-1$
like a freezing quintessence model. The present value of
$\omega_{\rm de_0}=-0.98$ is in good agreement with the result of
WMAP7.

(iii) The variation of the deceleration parameter $q$ shows that the
universe transits from an early matter dominant epoch, i.e. $q=0.5$,
to the de Sitter era, i.e. $q=-1$, in the future, as expected. The
accelerating expansion begins at $z_{\rm t}=0.74$, which is earlier
than that of the $\Lambda$CDM model. The current value of
$q_0=-0.56$ is in good accordance with the constraint coming from
the cosmography.

(iv) The evolutions of the dimensionless density parameters of DM
and DE clear that $\Omega_{\rm dm}$ decreases and $\Omega_{\rm de}$
increases during history of the universe. The present values of
dimensionless DM and DE densities are obtained as $\Omega_{\rm
dm_0}=0.229$ and $\Omega_{\rm de_0}=0.725$ which are in exact
agreement, surprisingly, with the results of WMAP7. Also the
obtained present density ratio $\Omega_{\rm de_0}/\Omega_{\rm
dm_0}=3.166\simeq{\mathcal O}(1)$ shows that in contrary to the
$\Lambda$CDM model, the cosmic coincidence problem is solved
naturally in the polytropic gas scenario.


\newpage
\begin{table}\small
\centering \caption{The observational $H(z)$ data
\cite{Hdata12,Hdata3}.}
\begin{tabular}{lcccccccccccc}\hline\hline\noalign{\smallskip}
$z$ & $0.09$ & $0.17$  & $\mathbf{0.24}$ & $0.27$ & $\mathbf{0.34}$
& $0.40$ & $\mathbf{0.43}$  & $0.88$ & $1.30$ & $1.43$ & $1.53$ &
$1.75$
\\\hline\noalign{\smallskip}
 $H(z)$ &  $69$ &$83$ & $\mathbf{79.69}$ & $70$ & $\mathbf{83.80}$ & $87$ &
$\mathbf{86.45}$  &$117$& $168$ & $177$ & $140$ & $202$
\\$1\sigma$ & $\pm12$ &$\pm8.3$ & $\mathbf{\pm2.32}$ & $\pm14$ &
$\mathbf{\pm2.96}$ & $\pm17.4$ & $\mathbf{\pm3.27}$ & $\pm23.4$ &
$\pm13.4$ & $\pm14.2$ & $\pm14$ &$\pm 40.4$
\\
\hline\hline
\end{tabular}
\label{Hdata}\\
\end{table}
\begin{table}
\centering\caption[]{The best fit values of the parameters
$\tilde{K}$ and $n$ within the 68.3\% ($1\sigma$) and 95.4\%
($2\sigma$) CLs for each observational data set for the polytropic
gas model. Columns 5, 6, and 7 show the current EoS parameter of DE,
the current deceleration parameter of the universe and the
transition redshift, respectively. The last row shows the best fit
result of the $\Lambda$CDM model using the full data sets for
comparison.}
\begin{tabular}{lcccccc}\hline\noalign{\smallskip}
{\rm Data} & $\tilde{K}$ &$n$ &$\chi^{2}_{\rm min}$&$\omega_{\rm
de_0}$&$q_{0}$&$z_{\rm t}$
\\\hline\hline\noalign{\smallskip}
 $\rm SN$ & $0.758_{-0.034-0.063}^{+0.031+0.055}$ & $-1.04$ &
$562.225$&$-0.98$&$-0.59$&$0.77$\\\\
$\rm SN+BAO$ & $0.760_{-0.031-0.063}^{+0.030+0.057}$ & $-1.01_{-0.12-0.25}^{+0.11+0.21}$ & $562.229$ &$-1$& $-0.59$ & $0.75$ \\\\
$$\rm SN+BAO\\+CMB$$ & $0.754_{-0.028-0.057}^{+0.028+0.054}$ &
$-1.03_{-0.09-0.18}^{+0.08+0.15}$ &
$562.366$&$ -0.98 $&$-0.58$&$0.75$\\\\
$$SN+BAO+\\CMB+OHD$$ & $0.742^{+0.024+0.048}_{-0.024-0.049}$ &
$-1.05^{+0.08+0.15}_{-0.08-0.16}$ &
$573.089$&$-0.98$&$-0.56$&$0.74$\\\\
$\Lambda\rm CDM$ & $-$ & $-$ & $573.552$ & $-1$&$-0.58$ & $0.72$\\
\hline
\end{tabular}
\label{best-fit}
\end{table}
\clearpage
\clearpage
\clearpage

\end{document}